\documentclass{appolb}
\usepackage{graphicx}
\usepackage{cite}
\usepackage{authblk}
\usepackage[T1]{fontenc}

\begin{document}
\title{Commissioning of the  \mbox{J-PET} detector for studies of decays of positronium atoms%
}
\author{
E.~Czerwi\'nski$^{a}$,
K.~Dulski$^{a}$, 
P.~Bia\l as$^{a}$,
C.~Curceanu$^{b}$,
A.~Gajos$^{a}$,
B.~G\l owacz$^{a}$,
M.~Gorgol$^{c}$,
B.~C.~Hiesmayr$^{d}$,
B.~Jasi\'nska$^{c}$,
D.~Kisielewska$^{a}$,
G.~Korcyl$^{a}$,
P.~Kowalski$^{e}$,
T.~Kozik$^{a}$,
N.~Krawczyk$^{a}$,
W.~Krzemie\'n$^{f}$,
E.~Kubicz$^{a}$,
M.~Mohammed$^{a,g}$,
Sz.~Nied\'zwiecki$^{a}$,
M.~Pa\l ka$^{a}$, 
M.~Pawlik-Nied\'zwiecka$^{a}$,
L.~Raczy\'nski$^{e}$,
Z.~Rudy$^{a}$,
N.G.~Sharma$^{a}$, 
S.~Sharma$^{a}$, 
R.Y.~Shopa$^{e}$, 
M.~Silarski$^{a}$,
M.~Skurzok$^{a}$,
A.~Wieczorek$^{a}$,
W.~Wi\'slicki$^{e}$,
B.~Zgardzi\'nska$^{c}$, 
M.~Zieli\'nski$^{a}$
P.~Moskal$^{a}$,
}

\affil{
       $^{a}$Faculty of Physics, Astronomy and Applied Computer Science, Jagiellonian University, 30-348 Cracow, Poland\\
       $^{b}$INFN, Laboratori Nazionali di Frascati, CP 13, Via E. Fermi 40, I-00044, Frascati, Italy\\
       $^{c}$Department of Nuclear Methods, Institute of Physics, Maria Curie-Sk\l odowska University, 20-031 Lublin, Poland\\
       $^{d}$Faculty of Physics, University of Vienna, Boltzmanngasse 5, 1090 Vienna, Austria\\
       $^{e}$\'Swierk Computing Center, National Centre for Nuclear Research, 05-400 Otwock-\'Swierk, Poland\\
       $^{f}$High Energy Physics Division, National Center for Nuclear Research, 05-400 Otwock-\'Swierk, Poland\\
       $^{g}$Department of Physics, College of Education for Pure Sciences, University of Mosul, Mosul, Iraq\\
     }

\maketitle
\begin{abstract}
The Jagiellonian Positron Emission Tomograph (\mbox{J-PET})
is a detector for medical imaging of the whole human body as well as for physics studies
involving detection of electron-positron annihilation into photons.

\mbox{J-PET} has high angular and time resolution and allows
for measurement of spin of the positronium and the momenta and polarization vectors of annihilation quanta.
In this article, we present the potential of the \mbox{\mbox{J-PET}} system for background rejection
in the decays of positronium atoms.
\end{abstract}
\section{Introduction}
Discrete symmetries C (charge conjugation), P (parity) and T (time reversal) and their combinations are
subject of vigorous investigation in various systems, like
nuclei~\cite{tric} or mesons~\cite{m28}. This interest is motivated among others by matter and anti-matter abundance asymmetry in the Universe (CP symmetry),
Lorenz invariance, unitarity and locality of quantum field theory (CPT symmetry) and uniqueness of time itself (T symmetry).
Tests of these symmetries are also performed in lepton systems. 
At present the hypothesis of CP symmetry conservation is excluded at 90\% confidence level in the accelerator neutrino oscillations
and should reach sensitivity greater than $3\sigma$ by 2026~\cite{t2k}.
Decays of positronium (a bound state of electron and positron) were investigated in search of CP and CPT violation~\cite{m47,m35} resulting in upper limits at the level of $10^{-3}$.

The Jagiellonian Positron Emission Tomograph (\mbox{J-PET}) system will, apart from medical applications
contribute to studies of discrete symmetries in decays of positronium~\cite{MoskalActa}.
Search of forbidden decays of the positronium triplet state, so-called ortho-positronium $\textrm{o-Ps}\to4\gamma$ and singlet state, so-called para-positronium
$\textrm{p-Ps}\to3\gamma$ will test the C symmetry.
Tests of the other fundamental symmetries and their combinations will be performed by the measurement of the 
expectation values of symmetry-odd operators constructed using $\textrm{o-Ps}$ spin ($\vec{S}$),
momentum of annihilation quantum ($\vec{k}_i$) and its polarization ($\vec{\epsilon}_i$).
Table~\ref{tab1} contains a list of such operators.
It is worth to mention that operators constructed with photons polarization vectors
are also available at \mbox{J-PET} system.
\begin{table}[h!]
\centering
\begin{tabular}{|c|c|c|c|c|c|}
\hline
 \textbf{Operator} & \textbf{C} & \textbf{P} & \textbf{T} & \textbf{CP} & \textbf{CPT} \\\hline
 $\vec{S}   \cdot \vec{k}_1$                                               & $+$ & $-$ & $+$ & $-$ & $-$   \\\hline
 $\vec{S}   \cdot (\vec{k}_1 \times \vec{k}_2$)                            & $+$ & $+$ & $-$ & $+$ & $-$   \\\hline
 $(\vec{S}  \cdot \vec{k}_1) (\vec{S} \cdot (\vec{k}_1 \times \vec{k}_2))$ & $+$ & $-$ & $-$ & $-$ & $+$   \\\hline
 $\vec{k}_1 \cdot \vec{\epsilon}_1$                                        & $+$ & $-$ & $-$ & $-$ & $+$   \\\hline
 $\vec{S}   \cdot  \vec{\epsilon}_1$                                       & $+$ & $+$ & $-$ & $+$ & $-$   \\\hline
 $\vec{S}   \cdot (\vec{k}_2 \times \vec{\epsilon}_1)$                     & $+$ & $-$ & $+$ & $-$ & $-$   \\\hline
\end{tabular}
\caption{
Discrete symmetries test operators for the $\textrm{o-Ps}\to3\gamma$ process~\cite{MoskalActa}.
The odd-symmetric operators are marked with "-" and are avaiable for studies at the \mbox{J-PET} system.
\label{tab1}
}
\end{table}
\section{Measurement technique}
The \mbox{J-PET} detector consists of 192 plastic scintillator strips ($500\times19\times7$~mm$^3$ each) made of EJ-230 organized in layers forming three cylinders (48 modules on radius 425~mm, 48 modules on radius 467.5~mm and 96 modules on radius 575~mm)~\cite{SzymonThese}.
Each scintillator at both of its ends is equipped with a R9800 Hamamatsu photomultiplier.
In the first test a point-like $^{22}\textrm{Na}$ source was placed in the center
and was covered by the XAD-4 porous polymer~\cite{xad4}. A test with the aluminium cylinder surrounding the source was also performed.\footnote{For
the future experiments the inner wall of the cylinder will be covered by the porous material presently under developement at the Maria Curie-Sklodowska University in Lublin.}
The positrons emitted from $^{22}\textrm{Na}\to^{22}\textrm{Ne}^*+e^++\bar{\nu}_e$ and
$^{22}\textrm{Ne}^*\to^{22}\textrm{Ne}+\gamma$ reaction chain with the $\gamma$ carrying 1274~keV
of energy
are stopped in the cylinder wall producing o-Ps
which may annihilate
emitting $3\gamma$. Detection of prompt photons with 1274~keV energy
from Ne deexcitation can be used as a start signal for the positronium lifetime measurement.
Gamma quanta from o-Ps annihilation are registered by means of Compton scattering inside the scintillator strips.
Two coordinates of the interaction point are determined
from the known strip position, while the position along the module is reconstructed from the time difference of signals registered by two photomultipliers
connected to the ends of that scintillator~\cite{J-NIM2014,J-1NIM2015,J-2NIM2014,J-2NIM2015,J-PMB2017}.
Annihilation point and time of o-Ps is then reconstructed using the trilateration technique~\cite{gps} taking advantage of the coplanarity of $\vec{k}_i$.
The known location of positron emission with reconstructed annihilation place allows for straightforward velocity ($\vec{v}$) direction determination.
The emitted positrons are longitudinally polarized $(\vec{P}=\frac{\vec{v}}{c})$ due to the parity violation in the $\beta$-decay.
Since the polarization of positron is to large extent preserved during the thermalization process~\cite{m48a,m48}, the known spin direction of positron provides the determination of o-Ps spin.
Gamma quanta undergo Compton scattering at \mbox{J-PET} plastic scintillators.  
Since Compton scattering is most likely to occur in the plane perpendicular to the electric vector of the photon~\cite{KleinNishina,Evans},
registration of annihilation and scattered quanta pairs allows to reconstruction of its linear polarization $\vec{\epsilon_i} = \vec{k_i} \times \vec{k^\prime_i}$,
where $\vec{k_i}$ and $\vec{k^\prime_i}$ denote momentum vectors of i-th gamma quantum before and after the Compton scattering, respectively~\cite{MoskalActa}.
For annihilation into $3\gamma$ with known $\vec{k}_i$ it is possible to calculate the energy loss of each $\gamma$ as described in Ref.~\cite{jpetEPJ}.
The energies of gamma quanta can be also estimated using independent method.
At the \mbox{J-PET} system the electric signals from photomultipliers are probed in time domain at 4 different amplitude thresholds resulting in
up to 8 time measurements (leading and trailing edge of the signal at each threshold)~\cite{jpetJINST,jpetDAQ}. Measured Time-Over-Threshold (TOT) value depends on
the amount of light registered by photomultiplier, while number of photons in the scintillator depends on the energy deposited by gamma quantum.
This method is especially useful to distinguish between prompt (E$_\gamma=1274$keV) and annihilation quanta (E$_\gamma\leq511$keV).
\section{Selection of data}
\begin{figure}[htb]
\centerline{%
\includegraphics[width=1.0\textwidth]{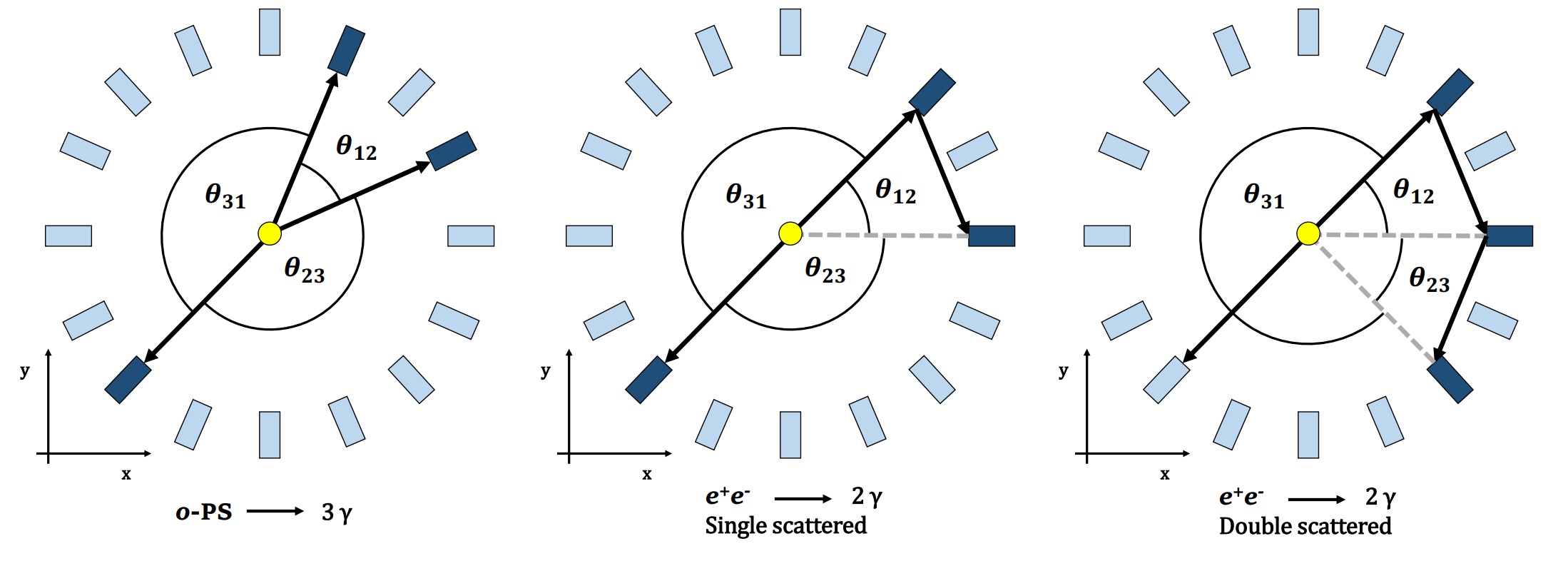}}
\caption{Definition of angles in XY view between registered gamma quanta with indexing fulfilling the condition $\theta_{12}\leq\theta_{23}\leq\theta_{13}$.
Light rectangles represent scintillators, dark rectangles - scintillators with registered hit and
yellow circle in the middle - annihilation region.}
\label{AngleScheme}
\end{figure}

Three data-taking campaigns are already concluded with the \mbox{J-PET} experiment for commissioning and physics studies which is equivalent to approx. 700~h
of data taking with different configuration and settings of detection system.\footnote{The fourth campaign is ongoing.}
Apart of threshold synchronization and calibration measurements the data were collected also for performance studies according to NEMA standards
which were so far evaluated based on the simulations only~\cite{KowalskiActa2015,KowalskiActa2016}.
Measurements were performed also with XAD-4 polymer target and 6 different polymer samples.
Data is acquired in the triggerless mode therefore an offline
reconstruction and signal selection is required for background rejection~\cite{jpetJINST,jpetDAQ,jpetFrame}.
For o-Ps data sample main background contribution is detector based scattering when 2 annihilation gamma quanta from p-Ps are registered
together with a single scattering or 1 annihilation $\gamma$ from p-Ps undergoes 2 scatterings are registered,
as schematically presented in Fig.~\ref{AngleScheme}.

\begin{figure}[htb]
\centerline{%
\includegraphics[width=0.60\textwidth]{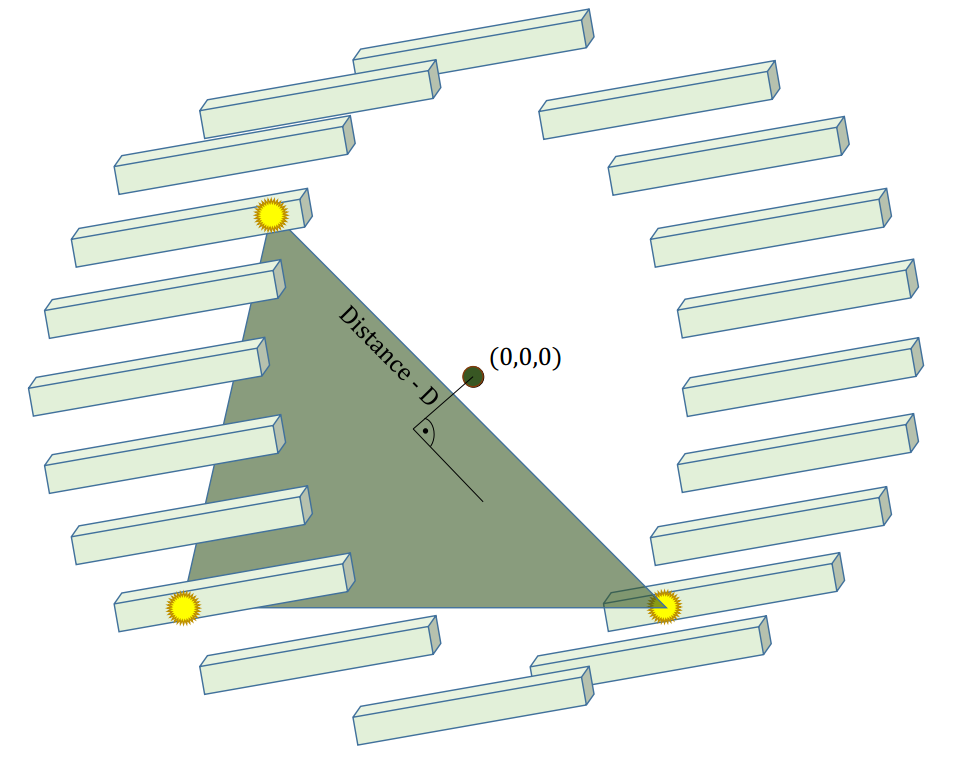}}
\caption{Distance between plane constructed with 3 hits and the source placed at the center of the detector.}
\label{DistanceScheme}
\end{figure}

These scattered events can be rejected in the process of the offline data reconstruction~\cite{jpetFrame}.
Here we present the preliminary results obtained from the second \mbox{J-PET} data-taking campaign for a 1~h~40~min measurement with the XAD-4 target of radius
$R\approx1$~cm
with only two following criteria applied offline:
\begin{enumerate}
\item
Number of registered hits in 200~ns time window equal to 3.
\item
TOT values for 2 hits correspond to annihilation quantum energy loss and TOT value for 3$^{rd}$ hit corresponds to prompt energy loss.
\end{enumerate}
\begin{figure}[htb]
\centerline{%
\includegraphics[width=0.70\textwidth]{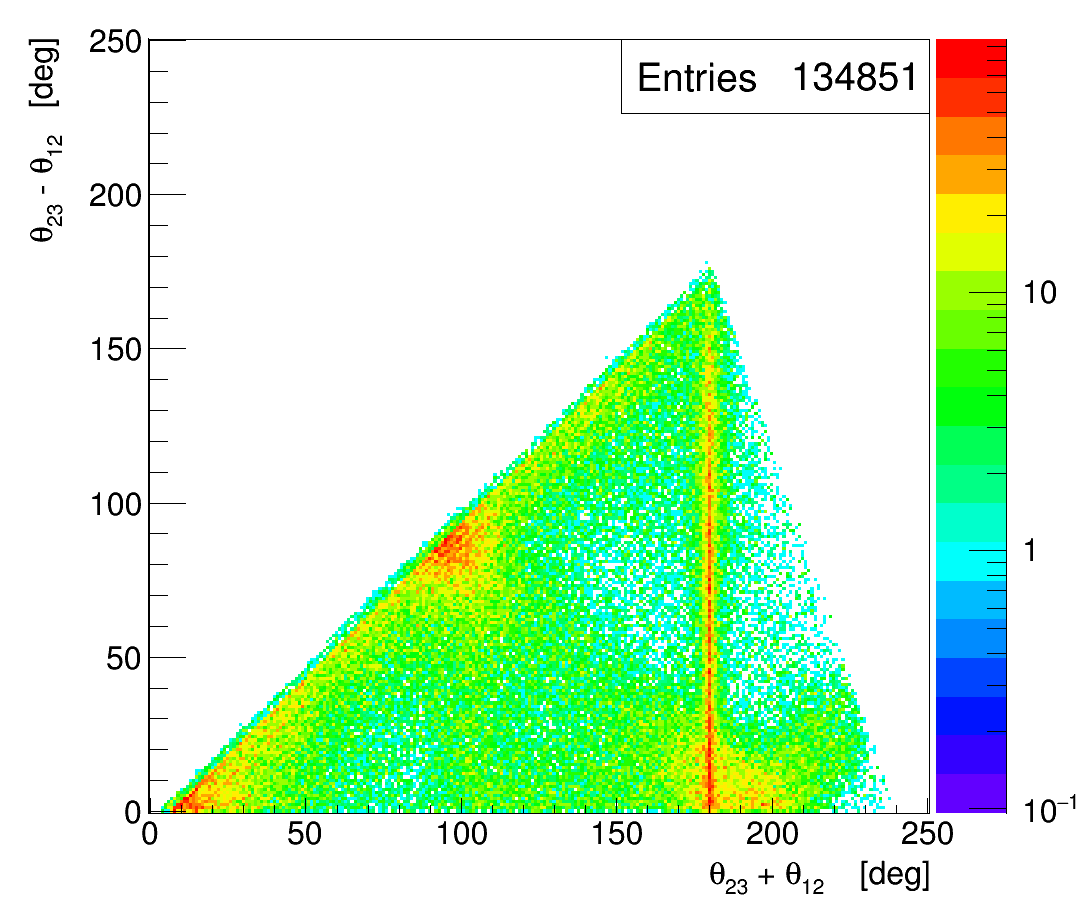}}
\caption{Distribution of registered events for $\theta_{23}+\theta_{12}$ and $\theta_{23}-\theta_{12}$ angles
as defined in Fig.~\ref{AngleScheme}. Statistics collected during 1~h~40~min of measurement.}
\label{AngleResult}
\end{figure}
The momentum vectors of all three gamma quanta, generated from o-Ps annihilation, lie in one plane.
Although the direction of prompt $\gamma$ emission is not correlated with such annihilation plane, in the XY view an o-Ps event fulfilling
the two above-mentioned criteria would have the same topology as left scheme in Fig.~\ref{AngleScheme} and therefore will occupy the region
of $\theta_{12}+\theta_{23}>180^{\circ}$ in Fig.~\ref{AngleResult}. The registered annihilation of p-Ps with single scattering
occurs as a line around $\theta_{12}+\theta_{23}\approx180^{\circ}$, while double scattered events are visible close to $(0^\circ,0^\circ)$.
The enhanced number of events close to $(100^\circ,100^\circ)$
correspond to scatterings by close to  $90^\circ$.
Registration of such scatterings is at most probable due to the geometry of the \mbox{J-PET} detector.
The conclusion from the distribution presented in Fig.~\ref{AngleResult} is that a significant part of the background can be rejected based on the angular correlations.
Another constraint to reject single scattered events is based on the plane constructed with three interaction points of gamma quanta with scintillators
as shown in Fig.~\ref{DistanceScheme}. For registered 3 annihilation $\gamma$ from o-Ps such plane should be close to annihilation place.
\begin{figure}[htb]
\centerline{%
\includegraphics[width=0.80\textwidth]{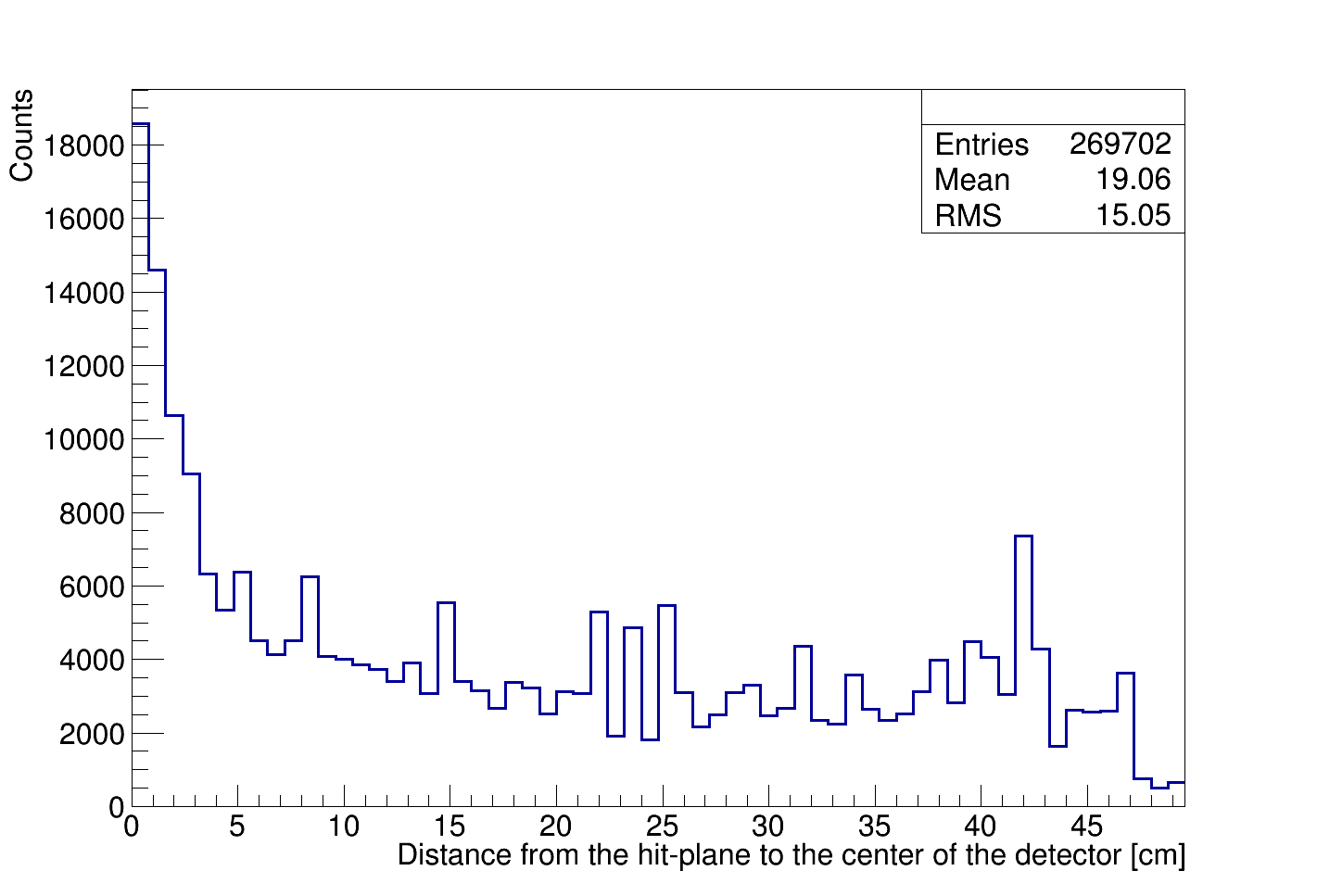}}
\caption{Distribution of the hit-plane distance from detector center. Statistics collected during 1~h~40~min of measurement.}
\label{DistanceResult}
\end{figure}
However, for a selection of events with two annihilation and one prompt photon
such plane would not be close to detector center, while p-Ps events with single scatterings would create a plane
crossing the center. Distribution of such distance is presented in Fig.~\ref{DistanceResult} with visible maximum at distances below about 5~cm.
The enhanced number of events for few bins of higher distance is due to the geometry of the detector.
\subsection{Summary}
The investigation of discrete symmetries is continuously performed in many systems.
So far no CP violation effects were discovered for the purely leptonic bound state such as positronium.
The Jagiellonian PET presently in commissioning stage will contribute to these studies.
The unique properties of the \mbox{J-PET} system are the usage of plastic scintillators for gamma quanta detection and capability of positronium spin determination
together with measurement of $\gamma$ polarization. High accuracy of time measurement together with rejection of background
based on the event geometry should allow for reaching a sensitivity of $10^{-5}$ for CP symmetry violation with the \mbox{J-PET} detector~\cite{MoskalActa}.
\section{Acknowledgments}
The authors acknowledge the technical support by A. Heczko, W. Migda\l\ and
the financial support from the Polish National Centre for Research and Development
through grant INNOTECH-K1/IN1/64/159174/NCBR/12, the EU
and MSHE Grant no. POIG.02.03.00-161 00-013/09, National Science Center Poland
based on decision number through Grant No. UMO-2016/21/B/ST2/01222.
B. C. Hiesmayr acknowledges gratefully the Austrian Science Fund FWF-
P26783.

\end{document}